\documentclass[prl,twocolumn,preprintnumbers,amsmath,amssymb,groupeaddress,superscriptaddress]{revtex4}

\usepackage[pdftex]{graphicx}
\usepackage{dcolumn}
\usepackage{bm}
\usepackage[latin1]{inputenc}

\newcommand{\ud}[1]{{#1^{\dagger}}}

\newcommand{\ket}[1]{\left| #1\right\rangle}

\newcommand{\mean}[1]{\langle#1\rangle}

\begin{document}


\title{Incoherent Mollow triplet}

\author{E. del Valle}
 \affiliation{School of Physics and Astronomy, University of Southampton, SO17 1BJ, Southampton, United Kingdom}%
\author{F. P. Laussy}
\affiliation{Walter Schottky Institut, Technische Universit\"at
M\"unchen, Am Coulombwall 3, 85748 Garching, Germany}

\date{\today}

\begin{abstract}
  A counterpart of the Mollow triplet (luminescence lineshape of a
  two-level system under coherent excitation) is obtained for the case
  of incoherent excitation in a cavity.  Its analytical expression, in
  excellent agreement with numerical results, pinpoints analogies and
  differences between the conventional resonance fluorescence spectrum
  and its cavity QED analogue under incoherent excitation.
\end{abstract}

\maketitle

Mollow~\cite{mollow69a} discovered a striking type of spectral shape
in the resonance fluorescence problem, where an atom is irradiated by
a strong laser beam. The celebrated \emph{Mollow
  triplet}~\cite{loudon_book00a}, that results from transitions
between atomic states that are dressed by the coherent light field,
has since been a testbed of nonlinear optics. It stands as one of the
fundamental spectral shapes of light-matter interaction, maybe second
only to the Rabi doublet.  Although the Mollow triplet is rooted in
quantum physics and bears much quantum features itself, it arises from
a fully classical light field. Its Hamiltonian, in the rotating frame
of the laser and at resonance, simply reads
$H_L=\Omega_L(\sigma+\ud{\sigma})$, with $\Omega_L^2$ the laser
intensity and $\sigma$ the only quantum operator, namely, the
two-level system annihilation operator. Including the spontaneous
decay of the emitter, in the Lindblad form $\mathcal{L}_\sigma
(\rho)=(2\sigma\rho
\ud{\sigma}-\ud{\sigma}\sigma\rho-\rho\ud{\sigma}\sigma)$, leads to a
master equation
$\partial_t\rho=i[\rho,H_L]+\frac{\gamma_\sigma}{2}\mathcal{L}_\sigma
(\rho)$ from which one obtains the famous Mollow triplet lineshape:
\begin{multline}
  \label{eq:MonJun28000330MSD2010}
   S_\mathrm{coh}(\omega)=\frac{\gamma_\sigma^2}{8 \Omega_L^2+\gamma_\sigma^2}
\delta(\omega)+\frac{1}{2\pi}\frac{\frac{\gamma_\sigma}{2}}{\big(\frac{\gamma_\sigma}{2}\big)^2+\omega^2}\\
 +\frac{\gamma_\sigma}{\pi}\frac{16\Omega_L^2-\gamma_\sigma^2-\omega^2}{\gamma_\sigma^4+4(\omega^2-4\Omega_L^2)^2+\gamma_\sigma^2(5\omega^2+16\Omega_L^2)} \,.
\end{multline}
It is composed of an elastic scattering peak, the Dirac
$\delta$~function, and the triplet itself, with a central Lorentzian
peak of full width at half maximum (FWHM) $\gamma_\sigma$ and two
satellite peaks at the symmetric positions $\pm \Re(R_L)$ with Mollow
splitting $R_L=\sqrt{(2\Omega_L)^2-(\gamma_\sigma/4)^2}$ and FWHMs
$3\gamma_\sigma/2$. This structure was observed a long time ago with
atoms~\cite{wu75a} and more recently also in a variety of solid state
systems~\cite{muller07a,ates09b,flagg09a,vamivakas09a,astafiev10a},
with, as befits the above description, coherent excitation.

In this text we consider a close counterpart of this fundamental
system, where the light field is initially fully quantized, and
becomes continuous as a result of an incoherent and continuous pumping
that feeds the system with a very large number of photons. This
situation is realized---as for quantization of the light field---in
cavity QED~\cite{haroche_book06a}, where quanta of a trapped standing
waves (the photons) can be brought to interact with an isolated
emitter, and---as for the incoherent pumping---with semiconductor
microcavities~\cite{delvalle_book10a}, where excitations are
continuously poured into the system with no external coherence fed in
by a driving field. The role of the emitter is, in this case, played
by a quantum dot placed in the antinode of the microcavity field. In
the cavity QED version of the resonance fluorescence physics, the
system is described by the Jaynes-Cummings Hamiltonian (still at
resonance): $H=g(\ud{a}\sigma+a\ud{\sigma})$ with the cavity mode also
quantized through the boson operator $a$. Cavity decay $\gamma_a$ as
well as incoherent pumping $P_\sigma$ are described like before with a
master equation
$\partial_t\rho=i[\rho,H]+\frac{\gamma_a}{2}\mathcal{L}_a(\rho)+
\frac{P_\sigma}2\mathcal{L}_{\ud{\sigma}}(\rho)$ where~$\rho$ is now
the density matrix for the combined two-level-emitter/cavity
system. We assume that $\gamma_\sigma\ll P_\sigma$, so much so that we
can neglect it for simplicity (it is already very small in a typical
system and we consider the case of large pumping). We also assume the
condition of strong-coupling, where $g\gg\gamma_a/4$. This regime was
recently observed in the cavity emission as a transition from the Rabi
doublet to a single lasing line~\cite{nomura10a}. Since this
transition has been claimed remaining in strong-coupling, it results
from climbing the Jaynes-Cummings ladder~\cite{delvalle09a}, and as
such, is a successful realization of quantum nonlinearities in these
systems. The importance of this breakthrough for microcavity QED is
however hindered by such a simple manifestation, in particular since
other mechanisms can also result in a similar behaviour of Rabi
splitting collapse without entering the quantum nonlinear
regime~\cite{munch09a}. 
%
Quantum features are generally better observed when probing the
quantum emitter, rather than the cavity, whose close connections with
the classical oscillator tend to surface rapidly and dominate
strongly. The theoretical description, which is straightforward in the
low excitation regime even when solving the system
exactly~\cite{delvalle09a}, becomes computationally demanding when the
lasing regime is approached, but good approximations can be
sought~\cite{arXiv_poddubny10a}. In this text, we consider this
physics of the highly nonlinear regime of the microcavity-QED system,
that is, Jaynes-Cummings physics under a strong incoherent pumping. We
find that, in good systems by the standard of today, a new type of
Mollow triplet is obtained in the direct QD emission spectrum. It is a
close counterpart of the classical Mollow triplet where light is
described by a classical field~\cite{mollow69a}, whereas it is here
described as numerous quanta of the cavity mode. The coherence is
acquired through the strong-coupling with the dot, resulting in
striking variations from the case where it is provided by an external
laser. We now describe them analytically.

\begin{figure}[t] 
\centering 
\includegraphics[width=.75\linewidth]{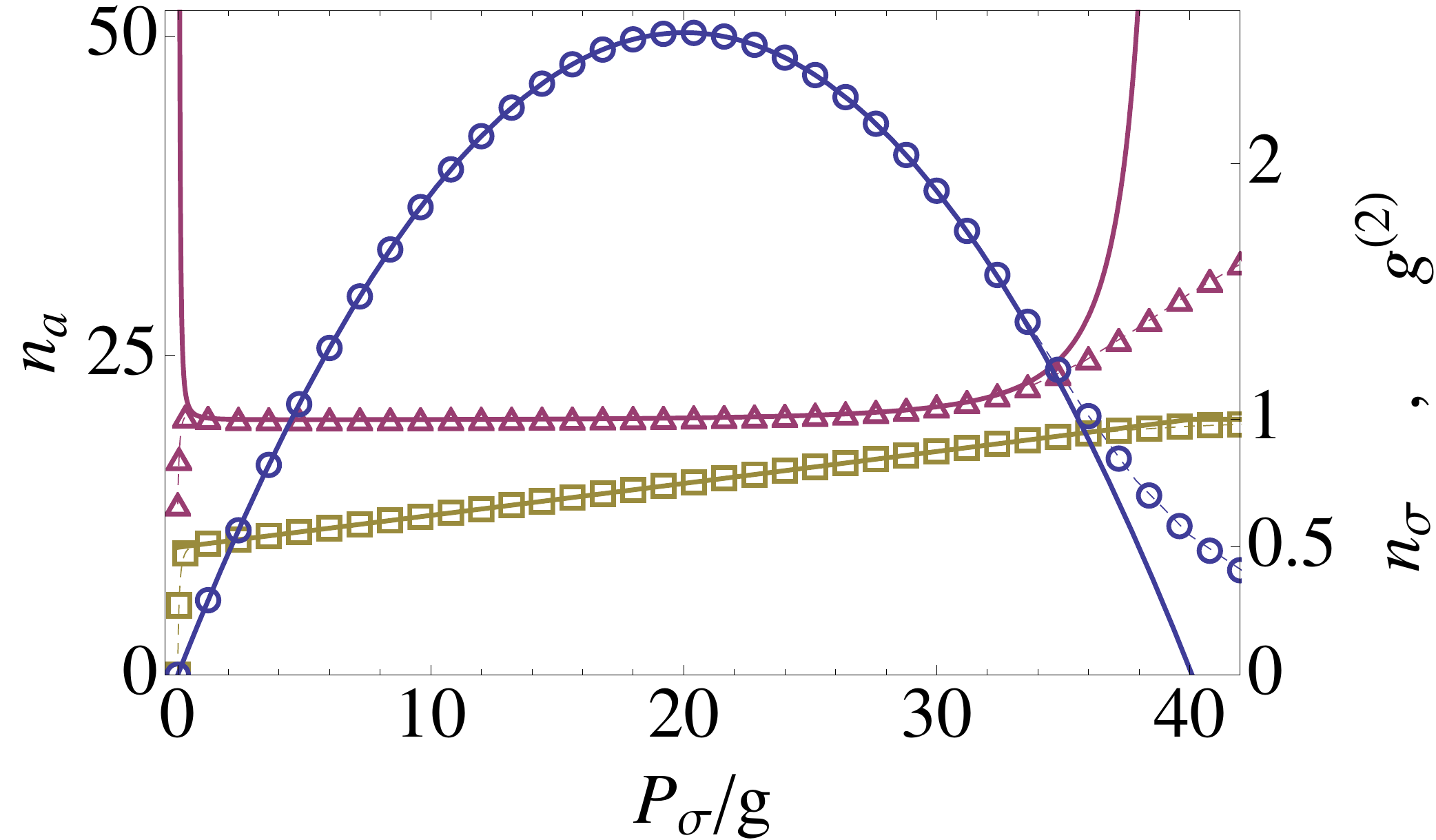}  
\caption{(Color online) Exact numerical solution (points) and their
  analytical approximation, Eqs.~(\ref{eq:SunJun27153112MSD2010}),
  (lines), for $n_a$ (blue/circles), $n_\sigma$ (brown/squares) and
  $g^{(2)}$ (purple/triangles), for $\gamma_a=0.1g$, as a function of
  pumping $P_\sigma/g$. The analytical solutions become unphysical
  when $P_\sigma=\kappa_\sigma$ (here at~$40g$), where $n_a=0$,
  $n_\sigma=1$ and $g^{(2)}$ diverges. They are very good
  approximations in the region of interest,
  Eq.~(\ref{eq:TueJul6130009CEST2010}).}
\label{fig:MoMar22195241WET2010}
\end{figure} 

\emph{Mollow regime} --- Whereas only one parameter (intensity) fully
describes the light in Mollow's description, the Jaynes-Cummings
picture requires from the start to take into account an infinite
number of correlators between the fields, that we can however relate
to each other~\cite{delvalle09a}:
$\mean{\ud{a}^na^{n-1}\sigma}=i\frac{\gamma_a}{2g}\mean{\ud{a}^na^n}$
and $\mean{\ud{a}^{n-1}a^{n-1}\ud{\sigma}\sigma}=\big[ P_\sigma
\mean{\ud{a}^{n-1}a^{n-1}}-\gamma_a \mean{\ud{a}^na^n}\big]/\big[
P_\sigma+\gamma_a(n-1)\big]$ (all others being zero in the steady
state).  From this follows a first relation for the populations of the
modes, $n_\sigma=\mean{\ud{\sigma}\sigma}$ and $n_a=\mean{\ud{a}a}$,
namely $n_\sigma=(P_\sigma -\gamma_a n_a)/P_\sigma$. This also allows
to obtain a self-contained equation for $\mean{\ud{a}^na^n}$:
\begin{equation}
  \label{eq:MoMar22173241WET2010}
  \mean{\ud{a}^na^n}=
  \frac{\frac{nP_\sigma}{P_\sigma+(n-1)\gamma_a}\mean{\ud{a}^{n-1}a^{n-1}}-\frac{2\gamma_a}{P_\sigma+n\gamma_a}\mean{\ud{a}^{n+1}a^{n+1}}}{1+\frac{P_\sigma+(2n-1)\gamma_a}{\kappa_\sigma}-\frac{2P_\sigma}{P_\sigma+n\gamma_a}
  +\frac{n\gamma_a}{P_\sigma+(n-1)\gamma_a}}\,,
\end{equation}
where $\kappa_\sigma=4g^2/\gamma_a$ is the Purcell rate of transfer of
population from the dot to the cavity mode. This recurrence equation
allows to compute $\mean{\ud{a}^na^n}$ for all $n$ as a function of
$n_a$ only. The solution for $n=0$ gives a good approximation for the
region where the cavity field behaves classically:
\begin{equation}
  \label{eq:SunJun27153112MSD2010}
  n_a\approx \frac{P_\sigma}{2\gamma_a}\Big(1-\frac{P_\sigma-\gamma_a}{\kappa_\sigma}\Big)\quad  \mathrm{and} \quad  n_\sigma\approx \frac{1}{2}\Big(1+\frac{P_\sigma-\gamma_a}{\kappa_\sigma}\Big)\,.
\end{equation}
The quality of this approximation is seen in
Fig.~\ref{fig:MoMar22195241WET2010} where it is compared with the
exact solution, computed numerically~\cite{delvalle09a}.  The second
order coherence function $g^{(2)}$ also admits a closed-form
expression (not given here but plotted in
Fig.~\ref{fig:MoMar22195241WET2010}) which is unity in good
approximation. The expressions Eqs.~(\ref{eq:SunJun27153112MSD2010})
for the populations have a clear physical meaning, that can be
discussed in terms of two parameters: the ``\emph{cavity feeding}'',
$F_a=P_\sigma/(2\gamma_a)$, and the ``\emph{dot feeding}'',
$F_\sigma=(P_\sigma-\gamma_a)/\kappa_\sigma$, efficiencies.  At low
pump, but still enough to be beyond the quantum
regime~\cite{delvalle09a}, i.e., $\gamma_a<P_\sigma\ll \kappa_\sigma$,
the cavity population increases linearly with pumping, with a half
occupied dot. This is the most effective region for accumulation of
photons in the cavity (the so-called \emph{one-atom
  laser}~\cite{mu92a}), with little disruption from incoherent
processes. At smaller efficiency~$F_\sigma$, the dot occupation also
increases linearly with pumping, quenching the linear increase of the
cavity population. $F_\sigma$ represents, therefore, the degree to
which the dot pumping succeeds in populating the dot itself, against
the coherent exchange of population that feeds the cavity with
efficiency $F_a$.  These expressions are thus valid until the dot
population is fully inverted, at $P_\mathrm{max}\approx\kappa_\sigma$,
then, the self-quenching dominates the dynamics, emptying the cavity
that goes to a thermal state.  The maximum population of the cavity,
$\max(n_a)\approx g^2/(2\gamma_a^2)$, is reached at the intermediate
rate $P_\sigma \approx \kappa_\sigma/2$. This identifies the regime of
interest for the observation of the Mollow triplet, where the cavity
field is intense ($n_a\gg 1$) and coherent (with a Poissonian photon
distribution, $T[n]=e^{-n_a}n_a^n/n!$ and $g^{(2)}=1$):
\begin{equation}
  \label{eq:TueJul6130009CEST2010}
  \gamma_a\ll g<P_\sigma<\kappa_\sigma\,.
\end{equation}

Now that we have a good and analytical description of the populations,
we turn to the optical emission spectrum, that we show can be obtained
in equally good approximations. The dot emission reads $n_\sigma\pi
S_\mathrm{inc}(\omega)\equiv
\Re\int_{0}^{\infty}\mean{\ud{\sigma}(0)\sigma(\tau)}e^{i\omega\tau}d\tau$. We
compute the two-time correlator $\mean{\ud{\sigma}(0)\sigma(\tau)}$ in
two steps: first, we solve the master equation in the steady state,
finding the density matrix elements $\rho_{m,i;n,j}$ (for $m$,
$n\in\mathbf{N}$ and~$i$, $j\in\{0,1\}$, photon and exciton indexes,
respectively). For the range of parameters of interest, we show that
they can be analytically expressed only in terms of the photon
distribution $T[n]$. Second, we apply the quantum regression
formula.

\emph{1. Steady state density matrix} --- We consider only elements
that are nonzero in the steady state: the populations
${p}_i[n]=\rho_{n,i;n,i}$ with $i=0,1$, corresponding to the
probability to have~$n$ photons with (${p}_1$) or without (${p}_0$)
exciton, and the off-diagonal terms ${q}_i[n]=\Im(\rho_{n,0;n-1,1})$,
corresponding to the coherence between the states~$\ket{n,0}$
and~$\ket{n-1,1}$. The master equation now reads:
\begin{subequations} 
\label{eq:MoMar22200554WET2010} 
\begin{align}
  \partial_t{p}_0[n+1]&=\mathcal{D}_\mathrm{phot}\{{p}_0[n+1]\}\\
  &-P_\sigma{p}_0[n+1]-2g\sqrt{n+1}{q}_\mathrm{i}[n+1]\,,\nonumber\\
  \partial_t{p}_1[n]&=\mathcal{D}_\mathrm{phot}\{{p}_1[n]\}\\
  &+P_\sigma
  (T[n]-{p}_1[n])+2g\sqrt{n+1}{q}_\mathrm{i}[n+1]\,,\nonumber\\
  \partial_t{q}_\mathrm{i}[n+1]&=\mathcal{D}_\mathrm{phot}\{{q}_\mathrm{i}[n+1]\}\\
  &-\frac{P_\sigma}{2}{q}_\mathrm{i}[n+1]+g\sqrt{n+1}({p}_0[n+1]-{p}_1[n])\,,\nonumber
\end{align} 
\end{subequations}
where we have separated the photonic dynamics into a superoperator
$\mathcal{D}_\mathrm{phot}$. Given that it is much slower than the dot
dynamics, one can solve the steady state ignoring
$\mathcal{D}_\mathrm{phot}$~\cite{scully_book02a}. The photon
distribution, $T[n]={p}_0[n]+{p}_1[n]$, remains unperturbed during the
excitation and interaction with the
dot. Equations~(\ref{eq:MoMar22200554WET2010}) then admit solutions in
terms of $T[n]$, i.e.,
${p}_0[n]\approx\frac{\kappa_a(n+1)}{P_\sigma+2\kappa_a(n+1)}T[n]$ and
${q}_\mathrm{i}[n]\approx\frac{-2g\sqrt{n}}{P_\sigma+2\kappa_an}T[n-1]$
where $\kappa_a=4g^2/P_\sigma$ is the Purcell rate of transfer of
population from the cavity to the dot. Our approximations of large
intensities imply $n+1\approx n$.

\emph{2. Two-time correlator and spectra} --- The two-time correlator
can be expressed as a sum
$\langle\ud{\sigma}(0)\sigma(\tau)\rangle=\sum_{n=0}^{\infty}Q[n](\tau)$,
where $Q[n]$ and other functions~$S_{0,1}[n]$ and~$V[n]$ are defined
through the quantum regression formula by coupled differential
equations ($n\geq 0$):
\begin{subequations} 
   \label{eq:TueMar23165009WET2010} 
   \begin{align}
     &\partial_\tau Q[n]=\mathcal{D}_\mathrm{phot}\{Q[n]\}\\
     &-\frac{P_\sigma}{2}Q[n]+ig(\sqrt{n}S_1[n]-\sqrt{n+1}S_0[n+1])\,,\nonumber\\
     &\partial_\tau S_0[n+1]=\mathcal{D}_\mathrm{phot}\{S_0[n+1]\}\\
     &-P_\sigma S_0[n+1]+ig(\sqrt{n}V[n+1]-\sqrt{n+1}Q[n])\,,\nonumber\\
     &\partial_\tau S_1[n]=\mathcal{D}_\mathrm{phot}\{S_1[n]\}\\
     &+P_\sigma (X[n]-S_1[n])-ig(\sqrt{n+1}V[n+1]-\sqrt{n}Q[n])\,,\nonumber\\
     &\partial_\tau V[n+1]=\mathcal{D}_\mathrm{phot}\{V[n+1]\}\\
     &-\frac{P_\sigma}{2}V[n+1]+ig(\sqrt{n}S_0[n+1]-\sqrt{n+1}S_1[n])\,.\nonumber
 \end{align} 
\end{subequations}
They are, like for the single-time dynamics, separated into a slow
photonic dynamics embedded in a superoperator
$\mathcal{D}_\mathrm{phot}$ that is $\tau$-independent in good
approximation, and a fast exciton and coupling dynamics that we can
solve analytically. Moreover, we have introduced the steady state
function $X[n]\equiv S_0[n](0)+S_1[n](0)$, in analogy with
$T[n]$. The initial conditions in
Eq.~(\ref{eq:TueMar23165009WET2010}), are the steady state values
$S_0[n+1](0)=i {q}_\mathrm{i}[n+1]$, $S_1[n](0)=0$,
$Q[n](0)={p}_1[n]$ and $V[n+1](0)=0$ (therefore, $X[n]=i
{q}_\mathrm{i}[n]$). After some long, but straightforward
algebra, we can find the expression for $Q[n](\tau)$ in terms of
${p}_{0,1}[n]$ and ${q}_i[n]$, which, in turn, are
expressed in terms of the statistics $T[n]$ (as shown
previously).  This allows to compute a closed-form solution for
$\langle\ud{\sigma}(0)\sigma(\tau)\rangle$, which is however lengthy
and not worth writing here. Its main physical features are to reveal
that each term in the sum over $n$, accounts for the 4 possible
transitions between the dressed states in the Jaynes-Cummings rungs
$n+1$ and $n$, as in the spontaneous emission case~\cite{delvalle09a}.
The first rung, or linear regime, is given by $n=0$ and consists of
only the two transitions of the Rabi doublet. Other rungs give rise to
a generalization of the Rabi frequency in the nonlinear regime, the
\emph{$n$th-rung inner} and \emph{outer Rabi frequencies}:
$R_{O,I}[n]=\sqrt{g^2(\sqrt{n+1}\pm\sqrt{n})^2-(P_\sigma/4)^2}$. In
the Mollow triplet regime ($P_\sigma>g$), all the peaks positioned at
the inner frequencies collapse at the centre (including the Rabi
doublet) giving rise to a single central peak.  Outer peaks remain
split at frequencies $\pm R_{O}[n]\approx
\pm\sqrt{4g^2n-(P_\sigma/4)^2}$.

\begin{figure}[t]
  \centering 
  \includegraphics[width=\linewidth]{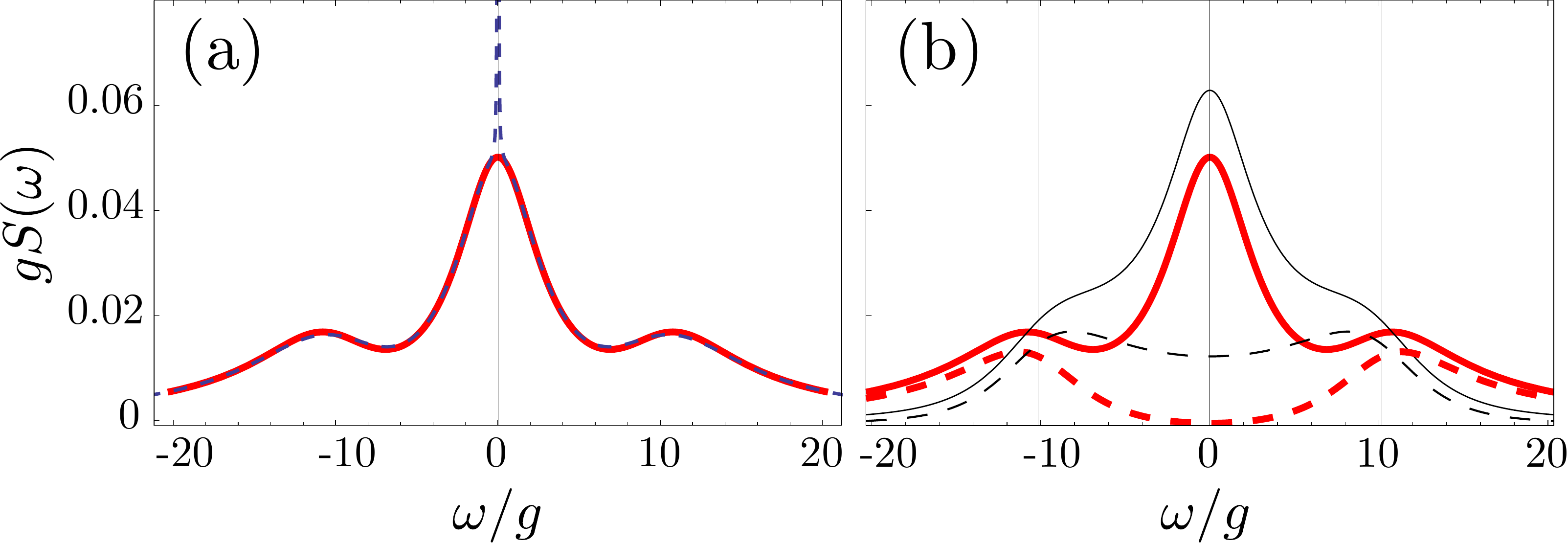}  
  \caption{(Color online) (a) Comparison between our analytical
    expression Eq.~(\ref{eq:MonJul5185141CEST2010}), without the
    elastic peak, (solid red) and the exact numerical solution (dashed
    blue).  (b) Difference between the incoherent (thick solid, red
    line) and the coherent (thin solid, black line) Mollow triplets,
    in equivalent conditions.  The satellite peaks, that cause the
    departure, are plotted in dashed. Parameters: $P_\sigma=6.3g$ and
    $\gamma_a=0.1g$.}
  \label{fig:MonJul5221814CEST2010}
\end{figure}

The spectrum obtained with the previous derivations can be further
simplified for the range of parameters in
Eq.~(\ref{eq:TueJul6130009CEST2010}), to give a compact analytical
expression. First, one considers only the coefficients with leading
terms in $n$, making use again of $n+1\approx n$. Furthermore, due to
the Poissonian statistics, only rungs with $n$ close to $n_a$
contribute significantly to the spectra allowing the substitution
$n\rightarrow n_a$ in $Q[n]$. The sum over $n$ simplifies thanks to
the normalization of the distribution function: $\sum_n
T[n]=1$. Finally, we neglect terms related to $\gamma_a$ before those
related to much larger rates, $P_\sigma$ and $\kappa_\sigma$, i.e., we
write the spectrum for these three rates only, through the
substitution $g^2=\kappa_\sigma\gamma_a/4$, and then simply set
$\gamma_a\rightarrow 0$. This results in the expression for
$S_\mathrm{inc}(\omega)$ in terms of $P_\sigma$ and $\kappa_\sigma$
only:
\begin{multline}
  \label{eq:MonJul5185141CEST2010}
  S_\mathrm{inc}(\omega)=\frac{P_\sigma}{\kappa_\sigma}\frac{\kappa_\sigma-P_\sigma}{\kappa_\sigma+P_\sigma}
  \delta(\omega)+\frac{1}{2\pi}\frac{\frac{P_\sigma}{2}}{\big(\frac{P_\sigma}{2}\big)^2+\omega^2}\\
  +\frac{1}{\pi}\frac{P_\sigma}{\kappa_\sigma+P_\sigma}\frac{(3\kappa_\sigma-P_\sigma)\omega^2-(\kappa_\sigma-3 P_\sigma)P_\sigma^2}{4\omega^4-P_\sigma(4\kappa_\sigma-9P_\sigma)\omega^2+\kappa_\sigma^2 P_\sigma^2}\,.
\end{multline}
This is our main result. The structure of the lineshape is the same as
that of its coherent counterpart,
Eq.~(\ref{eq:MonJun28000330MSD2010}): a Dirac $\delta$~function from
the elastically scattered laser light superimposed to a triplet. The
central Lorentzian peak has the same weight $1/2$ but with FWHM given
by the pump, $P_\sigma$. The two satellite peaks sit at $\pm \Re(R_O)$
with Mollow splitting:
\begin{equation}
  \label{eq:ThuJul8114120CEST2010}
  R_O=\frac{P_\sigma}{4}\sqrt{8 \kappa_\sigma /P_\sigma-9}
\end{equation}
and FWHM $3 P_\sigma/2$. The excellent agreement of our formula with
exact numerical results~\cite{loffler97a,delvalle09a} is shown in
Fig.~\ref{fig:MonJul5221814CEST2010}(a), where we superimpose in
dashed blue the numerical computation to, in solid red, the analytical
expression Eq.~(\ref{eq:MonJul5185141CEST2010}). Note the elastic peak
in the numeric as a very narrow central line.

\begin{figure}[t]
  \centering 
  \includegraphics[width=0.8\linewidth]{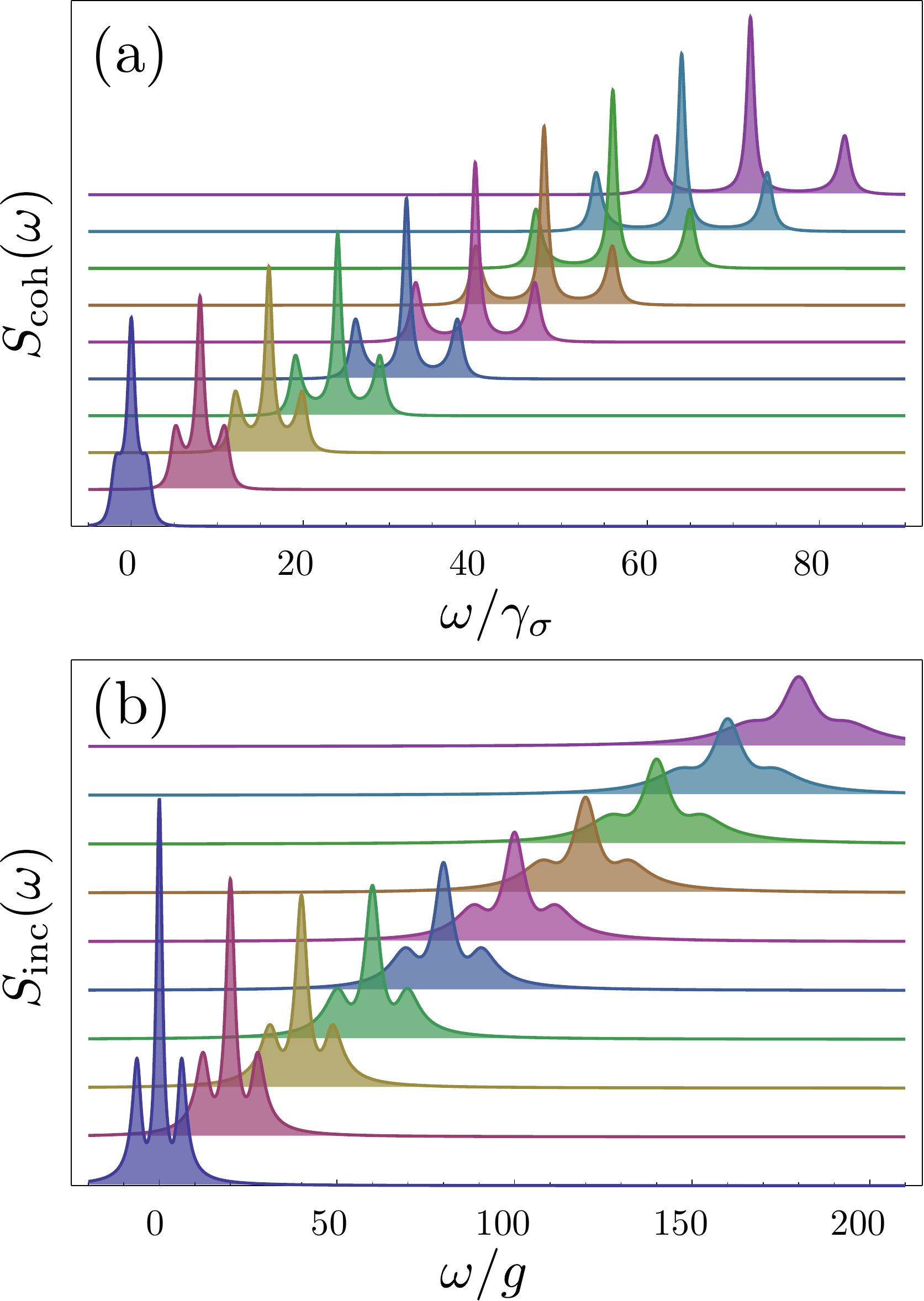}  
  \caption{(Color online) Evolution of the Mollow triplet when
    increasing (from bottom to top) (a) the coherent excitation
    ($\Omega_L/\gamma_\sigma$ from 1 to 5.5 by steps of 0.5) and (b)
    the incoherent excitation ($P_\sigma/g$ from 2 to 11 by steps
    of~1, with $\gamma_a=0.1g$).}
  \label{fig:MonJul5215129CEST2010}
\end{figure} 

Despite their similar structure, the lineshapes are intrinsically of a
different nature, as can be seen mathematically when reduced to their
simplest, dimensionless expression, where they depend only on one
parameter ($\Omega_L/\gamma_\sigma$ and
$P_\sigma/\kappa_\sigma$). Their similarities and differences can be
better appreciated when compared on physical grounds, when the laser
intensity for the conventional Mollow triplet,
Eq.~(\ref{eq:MonJun28000330MSD2010}), is taken the same as the average
population of the cavity under incoherent excitation (i.e., we take
$\Omega_L^2\rightarrow g^2 n_a\approx P_\sigma (\kappa_\sigma
-P_\sigma)/8$), and when the broadening of the dot includes
$P_\sigma$. In this way, we attempt the description of the incoherent
system with the theory of the coherent one. We obtain an expression
that shows the fundamental discrepancies between the two types of
triplets:
\begin{multline}
  \label{eq:MonJul5185736CEST2010}
  S_\mathrm{coh}(\omega)=\frac{P_\sigma}{\kappa_\sigma}\delta(\omega)+\frac{1}{2\pi}\frac{\frac{P_\sigma}{2}}{\big(\frac{P_\sigma}{2}\big)^2+\omega^2}\\
  -\frac{1}{\pi}\frac{P_\sigma\omega^2-(2\kappa_\sigma-3 P_\sigma)P_\sigma^2}{4\omega^4-P_\sigma(4\kappa_\sigma-9P_\sigma)\omega^2+\kappa_\sigma^2 P_\sigma^2}
  \,.
\end{multline}

Comparing this expression with Eq.~(\ref{eq:MonJul5185141CEST2010}),
one can see, indeed, that the central peak is the same in both cases,
as well as the position and broadening of the satellite peaks (third
terms have the same denominator), so the underlying structures bear
much similarities. However the satellite lineshapes differ
considerably, being strongly affected by the effect of incoherent
pumping, that renormalizes the rungs of the Jaynes-Cummings ladder,
and other factors such as the dot population, which under coherent
excitation shows opposite behaviour to that of
Eq.~(\ref{eq:SunJun27153112MSD2010}): $n_\sigma^\mathrm{coh}\approx
\frac{1}{2}(1-P_\sigma/\kappa_\sigma)$.  The shapes of these peaks are
shown in more details in Fig.~\ref{fig:MonJul5221814CEST2010}(b),
where they are plotted (in dashed) together with the whole triplets,
in the coherent (thin black) and the incoherent (thick red) cases.

Finally, Fig.~\ref{fig:MonJul5215129CEST2010} shows the natural
experimental configuration to demonstrate the new character of
nonlinear spectroscopy in microcavities under incoherent pumping, and
to contrast it with its coherent counterpart.  Increasing pumping, one
sees that in the coherent case (upper panel), the triplet is better
resolved, with a larger splitting, while in the incoherent case (lower
panel), the satellites overlap with the central line as a result from
pumping that splits them sublinearly,
Eq.~(\ref{eq:ThuJul8114120CEST2010}), and also increases their
broadening. The two phenomenologies, despite their deep
interconnections and common features, are strikingly different and the
evidence of the new one should pose no problem even on qualitative
grounds.


\bibliography{Sci,Elena,books,arXiv}

\end{document}